\documentclass{article}

\usepackage[final]{neurips_2021}

\usepackage[utf8]{inputenc} 
\usepackage[T1]{fontenc}    
\usepackage{hyperref}       
\usepackage{url}            
\usepackage{booktabs}       
\usepackage{amsfonts}       
\usepackage{nicefrac}       
\usepackage{microtype}      
\usepackage{xcolor}         
\usepackage{algorithm,algorithmicx,algpseudocode}
\usepackage{graphicx}
\usepackage[symbol]{footmisc}

\title{Towards dynamic multi-modal phenotyping using chest radiographs and physiological data}

\author{
  Nasir Hayat \\
  
  Engineering Division\\
  NYU Abu Dhabi, UAE \\
  \texttt{nh2218@nyu.edu}
  \\
  \And
  Krzysztof J. Geras  \\
   Department of Radiology\\
   NYU Grossman School of Medicine, USA \\
  \texttt{k.j.geras@nyu.edu}
   \And
    Farah E. Shamout  \\
   Engineering Division\\
   NYU Abu Dhabi, UAE \\
   \texttt{fs999@nyu.edu} \\
}

\begin{document}

\maketitle

\begin{abstract}
    The healthcare domain is characterized by heterogeneous data modalities, such as imaging and physiological data. In practice, the variety of medical data assists clinicians in decision-making. However, most of the current state-of-the-art deep learning models solely rely upon carefully curated data of a single modality.  In this paper, we propose a dynamic training approach to learn modality-specific data representations and to integrate auxiliary features, instead of solely relying on a single modality. Our preliminary experiments results for a patient phenotyping task using physiological data in MIMIC-IV \& chest radiographs in the MIMIC-CXR dataset show that our proposed approach achieves the highest area under the receiver operating characteristic curve (AUROC) (0.764 AUROC) compared to the performance of the benchmark method in previous work, which only used physiological data (0.740 AUROC). For a set of five recurring or chronic diseases with periodic acute episodes, including cardiac dysrhythmia, conduction disorders, and congestive heart failure, the AUROC improves from 0.747 to 0.798. This illustrates the benefit of leveraging the chest imaging modality in the phenotyping task and highlights the potential of multi-modal learning in medical applications.
  
\end{abstract}

\section{Introduction}  
Routine clinical practice generates and relies upon diverse data modalities, such as medical imaging, laboratory-test results, vital-sign measurements, textual data, or high-dimensional genomics data. Recent advancements in deep learning techniques have enabled the development of diagnostic and prognostic systems, which mainly rely on learning from a single data modality. However, less effort has been made to learn and integrate the heterogeneous data modalities representations to improve the performance of downstream tasks. 

The current paradigm of multi-modal learning in healthcare encompasses three data fusion strategies, which are broadly categorized into late, early, or joint fusion. \textit{Late fusion} performs an aggregation of information at the decision-level, such as by averaging predictions. One limitation of late fusion is that it does not learn a joint representation of the two modalities.
\textit{Early fusion} strategies combine features at the input-level, either in raw or processed format extracted via encoders, to compute the final prediction. For example, recent studies have used early fusion for prognosis and diagnosis of cancer \citep{cancer_survival, cancer_recurrence, Lung_Cancer, cancers_diag, Breast_Cancer,Therapy_Response}, by leveraging imaging, clinical, genomic, and molecular data. 
\textit{Joint fusion} strategies combine intermediate data representations of modality-specific encoders, then jointly fine-tunes the encoders. Through joint fusion, \cite{cervical} use cervigram imaging and relevant laboratory-test results for diagnosing cervical dysplasia, while \cite{carcinoma} fuse pre-processed computed tomography images with clinical data to diagnose hepatocellular carcinoma. 


%
Previous approaches commonly deal with semantically similar data modalities, i.e., the modalities represent the same set of labels with just structural differences. This limits their applicability to modalities that essentially do not predict the same set of labels, but rather provide rich clinical context that is helpful for downstream tasks. Another major limitation of these strategies is that they rely on the presence of both data modalities. This limits the training set to a subset of paired modalities and limits the applicability of such approaches in case of missing modalities at inference time.
%
\cite{missing} propose to reconstruct the features of the missing modality by conditioning on the available modality. However, it is not a trivial task to reconstruct the missing modality when dealing with data modalities that are not directly correlated, i.e., a biopsy report for skin tissue may not be able to reconstruct common thorax diseases features. Contrary to conventional multi-modal approaches in healthcare, the dynamic nature of our proposed training approach allows the integration of the auxiliary modality when it is available, which is more aligned with clinical practice.

Our contribution in this paper can be summarized as follows. We propose a novel dynamic approach towards integrating auxiliary data modalities. We jointly learn the data representations for the individual modalities and integrate the representations via a unified classifier. Since the approach does not rely on the availability of aligned data modalities, it inherently handles missingness and leverages all of the available modality-specific data.  

\begin{algorithm}[t!]
  \caption{The proposed dynamic training approach for chest X-rays (CXR) and electronic health records (EHR) data for the phenotyping task.}\label{alg}
  \begin{algorithmic}[1]
  \Require $\mathcal{X}^{cxr},\mathcal{X}^{ehr} , {Y}^{cxr}, Y^{ehr}$
      \For{$iter = 0$ to $N_{trainIter}$ }
        \State $\mathbf{x}^{cxr}_{b}$,  $\mathbf{y}^{cxr}_b  $ \Comment{Sample a CXR mini-batch with corresponding radiology labels}

        \State $\mathbf{p}^{cxr}_{b} = \mathbf{\phi}_{cxr} (\mathbf{\varphi}_{cxr}(\mathbf{x}^{cxr}_b))$ \Comment{Compute predictions for the mini-batch}
        \State $\mathbf{L}^{cxr}_{b} = \mathbf{L}_{cxr}(\mathbf{y}^{cxr}_{b}, \mathbf{p}^{cxr}_{b})$ \Comment{Compute loss using $\mathbf{p}^{cxr}_b $ and $\mathbf{y}^{cxr}_b $}
        
        \vspace{1mm}
        \State $\mathbf{x}^{ehr}_{b}$,  $\mathbf{y}^{ehr}_{b}$ \Comment{Sample an EHR mini-batch with corresponding phenotype labels}
        
        \State $\mathbf{p}^{cxr}_{b} = \mathbf{\phi}_{ehr} (\mathbf{\varphi}_{ehr}(\mathbf{x}^{ehr}_b))$ \Comment{Compute predictions for mini-batch}
        \State $\mathbf{L}^{ehr}_{b} = \mathbf{L}_{ehr}(\mathbf{y}^{ehr}_{b}, \mathbf{p}^{ehr}_{b})$ \Comment{Compute loss using $\mathbf{p}^{ehr}_b $ and $\mathbf{y}^{ehr}_b $}
        
        \State $\mathbf{L}_{b} = \mathbf{L}_{b}^{cxr} + \mathbf{L}_{b}^{ehr}$ \Comment{Update the encoders and classifiers}
        
        \vspace{1mm}
        \State $\mathbf{x}^{cxr}_{b}$, $\mathbf{x}^{ehr}_{b}$, $\mathbf{y}^{ehr}_{b}$ \Comment{Sample mini-batch with paired modalities \& phenotype labels}
        \State $ \mathbf{F}^{cat}_{b} = \mathbf{cat}(\mathbf{\varphi}_{cxr}(\mathbf{x}^{cxr}_b), \mathbf{\varphi}_{ehr}(\mathbf{x}^{ehr}_b)) $ \Comment{Concatenate the feature representations}
        \State $\mathbf{p}^{cat}_{b} = \mathbf{\phi}_{unified}(\mathbf{F}^{cat}_{b})$ \Comment{ Compute phenotype predictions from concatenated features}
        
        \State $\mathbf{L}^{cat}_{b} = \mathbf{L}_{cat}(\mathbf{y}^{ehr}_{b}, \mathbf{p}^{cat}_{b})$ \Comment{Update $\mathbf{\phi}_{unified}$ using loss $\mathbf{L}^{cat}_{b}$}
        
      \EndFor
  \end{algorithmic}
\end{algorithm}

\section{Method} Our proposed approach is designed to work for any two modalities and is flexible to integrate additional modalities. To explain the methodology, we consider two data modalities: $\mathcal{X}^{1}$ and $\mathcal{X}^{2}$. Let's assume the set $\mathcal{X}^{1}$ is associated with labels ${Y}^{1} = \{1,\cdots R\}$, and set $\mathcal{X}^{2}$ is associated with labels ${Y}^{2} = \{1,\cdots P\}$, where $R$ and $P$ denote the total number of labels, respectively. If ${Y}^{1}={Y}^{2}$, then the modalities are considered to be homogeneous as they are assigned the same labels. During a single training iteration, mini-batches from $\mathcal{X}^{1}$ and $\mathcal{X}^{2}$ are sampled. The mini-batches of the two modalities are not paired, i.e., a sample may have $\mathcal{X}^{1}$ but not $\mathcal{X}^{2}$. We extract the features of these samples from the penultimate layers of the defined encoders, $\mathbf{\varphi}_{1}$ and $\mathbf{\varphi}_{2}$, respectively. These features are then processed through respective classification modules $\mathbf{\phi}_{{1}}$ and $\mathbf{\phi}_{{2}}$ to compute the losses with respect to each modality's set of labels. Since each instance in the training set is associated with at least one binary label, then the loss functions are defined using binary cross-entropy (BCE). The encoders ($\mathbf{\varphi}_{1}$ and $\mathbf{\varphi}_{2}$) and classification modules ($\mathbf{\phi}_{{1}}$ and $\mathbf{\phi}_{{2}}$) are updated through these losses.

Then, we sample a mini-batch of paired samples, i.e. samples with both $\mathcal{X}^{1}$ and $\mathcal{X}^{2}$. After processing the modalities using their respective encoders, their feature vectors are concatenated into a single vector, which is then processed by a unified classification module. The unified module predicts the labels of interest in relation to the task at hand, which could be ${Y}^{1}$ or ${Y}^{2}$ if ${Y}^{1}\neq{Y}^{2}$. In this step, we only update the parameters of the classifier. At inference time, the method uses the modality-specific prediction of the task labels to mitigate missingness. For example, it would consider $\mathcal{X}^{2}$ as the base modality, if the task of the unified classifier is to predict ${Y}^{2}$, in the case that the auxiliary modality ($\mathcal{X}^{1}$) is missing. 

\begin{table}[t]
\parbox[t]{.49\linewidth}{
 \caption{\footnotesize{\textbf{Number of samples for each modality within each split.}
 } {CXR} represents the number of chest radiographs, {EHR} represents the total number of ICU stays with physiological data extracted from the electronic health records (EHR), and {Pairs} represents the number of ICU stays with both CXR and EHR data. \textbf{Bold:} The results are summarized for the test set of pairs.  } \vspace{-2mm}
    \centering
     \resizebox{0.7\linewidth}{!}{
    \begin{tabular}{c | c c c}
        \toprule 
        \bfseries Split & \bfseries CXR & \bfseries EHR & \bfseries Pairs \\
        \midrule
        Train & 325188 & 42628 & 8056 \\
        Val & 15282 & 4802 & 919 \\
        Test & 36625 & 11914 & \textbf{2240} \\

        
        \bottomrule
    \end{tabular}
     }
    \vspace{-1mm}
    \label{splits}
    
}
\hfill
\parbox[t]{.49\linewidth}{
    \caption{\footnotesize{\textbf{Performance results on the `Pairs' test set.}
 } The macro average AUROC is shown for \textbf{all}, \textbf{acute},  \textbf{mixed}, and \textbf{chronic} diseases. \textbf{EHR} uses only physiological data while two joint fusion strategies are used as baselines. \textbf{Unified} is our proposed approach.}
  \vspace{-5mm}
    \centering
     \resizebox{1.0\linewidth}{!}{
    \begin{tabular}[t]{c | c c c c}
        \toprule 
        \bfseries Modality & \bfseries all  & \bfseries acute & \bfseries mixed & \bfseries chronic \\
        \midrule
        
        EHR &  0.740 & 0.743 & 0.747 &  0.731 \\
        Joint I $\dag$ & 0.724 & 0.724 & 0.742 & 0.712 \\
        Joint II $\ddag$ & 0.756 & 0.746 & 0.788 & 0.753 \\
        \midrule
        Unified & \textbf{0.764} & \textbf{0.752} &  \textbf{0.798} & \textbf{0.759} \\
        
        \bottomrule
        \multicolumn{5}{p{7cm}}{\scriptsize{$\dag$ The penultimate representations of the encoders are concatenated to predict phenotype labels. The overall network is initialized randomly \& trained from scratch. }}\\ 
        \multicolumn{5}{p{7cm}}{\scriptsize{$\ddag$ The penultimate representations of the encoders are concatenated to predict phenotype labels, but the encoders are pre-trained with their respective labels.}}\\
    \end{tabular}
     }
    \vspace{-3mm}
    \label{results}
    \hspace{-2em}
    }
\end{table}
\section{Experiments and results}
We ran preliminary experiments for a phenotyping task, defined by \cite{benchhmark}, which aims to classify which of 25 acute care conditions are present in a given patient intensive care unit (ICU) stay. 
In the original work \citep{benchhmark}, the authors use data extracted from electronic health records (EHR) to predict the 25 conditions, where they conventionally select and preprocess a subset of 17 time-series physiological features from the MIMIC-III dataset \citep{mimic3, physio_challenge}. In this work, we extract the same physiological features with a modified pipeline for the updated database MIMIC-IV \citep{mimic4, physionet} to represent $\mathcal{X}^{ehr}$ as the first (base) modality. The features are discretized into time-series events within two-hour windows. We parameterize $\mathbf{\varphi}_{ehr}$ as a long short-term memory (LSTM) network \citep{lstm} to encode the time-series physiological features and $\mathbf{\phi}_{ehr}$ as a sequence of linear and sigmoid layers to predict the 25 phenotype labels.

For the second modality, we extract a set of chest radiographs from MIMIC-CXR \citep{mimiccxrjpg}. The chest X-rays (CXR) are considered as the auxiliary modality $\mathcal{X}^{cxr}$ and we parameterize the visual encoder $\mathbf{\varphi}_{cxr}$ as  ResNet-34 \citep{resnet34} and $\mathbf{\phi}_{cxr}$ as a sequence of linear and sigmoid layers to predict 14 radiology labels. To create the set of paired modalities, we matched the ICU stays with chest radiographs based on the patient identifier and the timestamp of collection. The unified classifier is set to predict the 25 phenotype labels. We outline the dynamic training pipeline in Alg. \ref{alg}. An overview of the cohort based on a random training, validation, and test split is shown in Table \ref{splits}.

Table \ref{results} summarizes the results for the test split where data from both modalities are available. We report the average AUROC across all 25 diseases, 12 critical diseases, 8 chronic diseases, and 5 “mixed” diseases as they are recurring or chronic with periodic acute episodes. Our unified approach achieves the best AUROC across all categories, with 0.764 AUROC overall, compared to 0.740 AUROC when using physiological data only. We also observe a similar pattern to that reported in \citep{benchhmark}, where chronic diseases are more difficult to predict than acute ones when relying on physiological data only. However, it is interesting to note that our unified approach improves the performance across both acute and chronic diseases, highlighting the importance of the chest radiographs in providing additional information for this phenotype classification task. 

\section{Conclusion} 
In this work, we propose a dynamic training approach to integrate multi-modal data. During training, our approach makes more optimal use of the data modalities to learn better representations and integrates features simultaneously. Our results show that the incorporation of the auxiliary imaging modality improved performance in the benchmark phenotyping task, which better mimics real-world clinical practice and can be considered a step towards improving general intelligence in the medical domain.


\section*{Potential negative societal impact}
Even though our work highlights the potential of multi-modal learning in healthcare, especially by fusing imaging and non-imaging modalities, the application is limited to chest radiographs and 17 physiological features. These modalities may not be available to patients in low-resource settings, deeming the work to be less accessible. Additionally, we focus on a set of 25 phenotype labels only, as in the benchmark work. This may also make the work less relevant for patients with uncommon or rare diseases. 

\bibliographystyle{plainnat}
\bibliography{sample}

        

        

\end{document}